# Asymmetric Viscothermal Acoustic Propagation and Implication on Flow Measurement

Yong Chen, Bo Yuan, Xiaoqian Chen, Lei Qi

*Abstract*—**In the application of high frequency acoustic flow measurement, viscothermal dissipation and asymmetric acoustic modes cannot be overlooked. Present paper mathematically formulates asymmetric linear disturbance dynamics in terms of velocity and temperature disturbances based on the conservations of mass, momentum and energy. An iterative calculation procedure, which is similar to Galerkin method, is presented. Numerical analysis of asymmetric acoustic features (phase velocity and attenuation coefficient) are comprehensively given under the effects of viscothermal dissipation and shear flow convection. In the end, flow measurement performance of asymmetric acoustic modes is literally discussed. Numerical study shows that viscothermal dissipation affects the cut-on frequency of acoustic modes and couples nonlinearly with shear convection when the flow Mach number is large. These parameters impose significant influences on measurement performance. Each acoustic mode has inherent measurement derivation which can be theoretically used to compensate the acoustic flow measurement error. Apparent prediction error may occur if the viscothermal dissipation is taken out of consideration.**

## I. INTRODUCTION

IN the application of acoustic pipeline flow measurement, accurate prediction of acoustic wave propagation is of great importance on improving measurement performance[1, 2]. Using the inviscid fluid model which neglects the viscothermal dissipation, Lechner [3] investigated measurement performance in the presence of an axially shear flow. Based on the same numerical method, Willatzen [4] made constructive comments on the work of Lechner. Continually, Willatzen [5] analysed the acoustic flow measurement using different isentropic acoustic models for inviscid fluid, which were Mungur and Plumblee [6] model, simplified irrotational model[7], and local-plane-wave(LPW) model[8], under three different shear flow profiles. The above mentioned contributions showed that different acoustic models, acoustic modes and flow profiles could lead to different performance predictions for flow measurement.

To get good detection of transit time, high frequency acoustic disturbance is preferred in the acoustic flow measurement, which, on the other hand, will cut on many axisymmetric and asymmetric modes propagating in the flow. To investigate the corresponding measurement performance, the authors[9] formulated the asymmetric isentropic acoustic dynamics on the basis of Mungur and Plumblee [6] model. Meanwhile, the acoustic impedance of the non-rigid wall was modelled and its influence on measurement performance was numerically studied.

The above mentioned contributions neglect the viscothermal influence which also affect the features of acoustic dynamics[10]. Importantly, in the application of acoustic flow measurement for a small-radius pipeline where high frequency acoustic waves are used, viscothermal dissipation becomes apparent and its influence should be considered[11]. Under the assumption of low frequency disturbance, simplified viscothermal acoustic dynamics has been mathematically established for shear flow[12-15] and uniform flow[16] in the literature. As the low frequency assumption simplifies the physical dynamics in the radial direction, the model cannot be applied to situations where higher modes other than plane mode exist[17].

If fluid thermal conduction is neglected, Elvira-Segura [18] gave mathematical model of axisymmetric viscous acoustic dynamics. Baik, et al. [19] extended to model the asymmetric viscous acoustic dynamics. Recently, Dokumaci [11] proposed a viscothermal acoustic model which can analyse both axisymmetric and asymmetric modes. However, the model is restricted to conditions where assumptions of uniform flow or stationary fluid can be accepted. Neglecting the acoustic dynamics in circumferential direction, the authors gave mathematical formulation of axisymmetric acoustic dynamics in the presence of shear[17] and uniform[20] flow profiles and analysed the corresponding measurement performance[21, 22].

As high frequency acoustic dynamics is extremely affected by the asymmetric vibrations and viscothermal dissipation, present paper tries to give mathematical formulation of asymmetric viscothermal acoustic dynamics in the presence of an axially shear flow. Such model is used to theoretically analyze the acoustic flow measurement performance. Theoretically, hydrodynamic disturbances also exist and propagate along with

Revised manuscript received by April 2018. This work was supported by the National Natural Science Foundation of China (91741107, 11504427, 61601489, and 51675525). The CAST-BISEE Foundation, named “on-orbit non-contact propellants’ multiple parameters measurement”, supports present work.

Yong Chen is with the National University of Defense Technology, China (email: literature_chen@nudt.edu.cn). He is also with the Department of Applied Mathematics and Theoretical Physics, University of Cambridge. (email: yc433@cam.ac.uk).

Bo Yuan is with the Department of Machinery and Electrical Engineering, Logistical Engineering University, Chongqing, China. (email: ramboyuanbo@aliyun.com).

Xiaoqian Chen is with the National Institute of Defense Technology Innovation, Academy of Military Science, Changsha, China. (email: chenxiaoqian@nudt.edu.cn).

Lei Qi is with the Beijing Institute of Spacecraft Environment Engineering, Beijing, 10094, China. (email:qilei511@126.com).

the flow as shown by Dokumaci [11] and Peat [12]. However, their features will be little interested in the applications of acoustic flow measurement as they are nearly convected by the mean flow profile. As a result, present paper omits theoretical analysis of hydrodynamic disturbances even the established model can cover such analysis.

The layout is as follows. Section 2 gives theoretical formulation of viscothermal disturbances in the presence of an arbitrary axially shear flow. Section 3 presents comprehensive analysis on features of asymmetric acoustic modes and performance of high frequency acoustic flow measurement. Section 4 concludes the work in this paper.

## II. MATHEMATICAL FORMULATIONS

This section deals with mathematical deductions of asymmetric acoustic waves due to the effects of viscothermal dissipation and shear flow convection. The disturbance is assumed to be linear and then first-Taylor's expansion method is preferred. Thus, one can get the following expressions for pressure, density, temperature, entropy, and velocity

$$p = p_0 + p', \rho = \rho_0 + \rho', T = T_0 + T', s = s_0 + s', \mathbf{v} = \mathbf{v}_0 + \mathbf{v}'. \quad (1)$$

In the above equation, the mean incompressible flow is subscripted by '0' meanwhile the linear acoustic dynamics is superscripted by a prime. In the subsonic flow where no external heats and forces are added, disturbances to the following parameters can be ignored: shear viscosity ($\eta$), bulk viscosity ($\zeta$), specific heat at constant pressure ($c_p$), thermal conductivity ($\kappa_{\mathrm{th}}$), coefficient of volume thermal expansion ($\beta$), and ratio of specific heat coefficient ($\gamma$). These parameters are taken to be constant in the subsonic flow in the further assumption.

### A. *Mathematical formulations*

As described before, the mean flow is assumed to be incompressible ($\rho_0 = \text{constant}$) and axially shear( $\mathbf{v_0} = \left[0, 0, v_0(r)\right]$ ) without external forces and heat additions, the governing equation of mean flow in the cylindrical circular coordinate, of which the radial, circumferential and axial coordinates are denoted by $r$ , $\theta$ and $z$ respectively, can be written by[17]

$$\nabla p_0 = \eta \nabla^2 \mathbf{v}_0, \quad (2)$$

$$\kappa_{\mathrm{th}} \nabla^2 T_0 + \eta \beta T_0 \left(\mathbf{v}_0 \cdot \nabla^2 \mathbf{v}_0\right) + \frac{\eta}{R^2}\left(\frac{\mathrm{d}v_0}{\mathrm{d}r}\right)^2 = 0. \quad (3)$$

In the above equation, the radial coordinate ($r \in [0,1]$) is normalized through the pipeline radius $R$ . Under the neglect of axial and circumferential gradients of the mean temperature, it satisfies

$$\frac{\kappa_{\mathrm{th}}}{\eta \bar{c}_0^2}\frac{1}{r}\frac{\mathrm{d}}{\mathrm{d}r}\left(r\frac{\mathrm{d}T_0}{\mathrm{d}r}\right) + \beta T_0 \frac{M\mathrm{d}}{r\mathrm{d}r}\left(r\frac{\mathrm{d}M}{\mathrm{d}r}\right) + \left(\frac{\mathrm{d}M}{\mathrm{d}r}\right)^2 = 0, \quad (4)$$

where the local Mach number $M(r) = v_0(r)/\bar{c}_0$ is introduced and $\bar{c}_0$ denotes the averaged adiabatic sound speed.

On the other hand, the governing equation of linear disturbance can read[17]

$$\frac{\mathrm{D}\rho'}{\mathrm{D}t} + \rho_0 \nabla\cdot\mathbf{v}' = 0, \quad (5)$$

$$\begin{aligned} &\frac{\mathrm{D}\mathbf{v}'}{\mathrm{D}t} + \left(\mathbf{v}'\cdot\nabla\right)\mathbf{v}_0 \\ &= -\frac{1}{\rho_0}\nabla p' + \frac{\eta}{\rho_0}\nabla^2\mathbf{v}' + \frac{1}{\rho_0}\left(\zeta + \frac{\eta}{3}\right)\nabla\left(\nabla\cdot\mathbf{v}'\right), \end{aligned} \quad (6)$$

$$\begin{aligned} &\rho_0 T_0 \left(\frac{\mathrm{D}s'}{\mathrm{D}t} + \mathbf{v}'\cdot\nabla s_0\right) + \rho_0\left(\mathbf{v}_0\cdot\nabla s_0\right)T' \\ &= \kappa_{\mathrm{th}}\nabla^2 T' + \frac{2\eta}{R}\frac{\mathrm{d}v_0}{\mathrm{d}r}\left(\frac{\partial v_z'}{R\partial r} + \frac{\partial v_r'}{\partial z}\right), \end{aligned} \quad (7)$$

where $\mathrm{D}/\mathrm{D}t = \partial/\partial t + \left(\mathbf{v}_0\cdot\nabla\right)$, the substantial derivative operator, is used. According to the thermodynamic relations[7, 11], the entropy and density disturbances can be written as

$$s' = \frac{c_p T'}{T_0} - \frac{\beta}{\rho_0}p', \rho' = \frac{\gamma}{c_0^2}p' - \beta\rho_0 T'. \quad (8)$$

Due to the variable mean temperature(Eq.(4)) along the radial coordinate, the adiabatic sound speed $c_0$ is a function of the radial coordinate( $r$ ).

Since wave-like solutions to Eqs. (5)-(7) are of interest, the dependences of perturbations on time, circumferential and axial coordinates can be written as $\exp\left[\mathrm{i}\left(\omega t - m\theta - k_0 K z\right)\right]$, where i , $\omega$, $m$ and $K$ are respectively the unit imaginary number, angular disturbance frequency, circumferential and axial wavenumbers nondimensionalized by the total wavenumber $k_0 = \omega/\bar{c}_0$ . The substantial derivative operator then can be written by $\mathrm{D}/\mathrm{D}t = \mathrm{i}\omega(1 - KM)$ . As a result, the pressure, velocity and temperature disturbances can be nondimensionalized by

$$\begin{aligned} &\left[v_r', v_\theta', v_z'\right] = \bar{c}_0\left[\varphi_r(r), \varphi_\theta(r), \varphi_z(r)\right], \\ &\left[p', T'\right] = \left[\bar{p}_0\varphi_p(r), \bar{T}_0\varphi_T(r)\right], \end{aligned} \quad (9)$$

where the dependence $\exp\left[\mathrm{i}\left(\omega t - m\theta - k_0 K z\right)\right]$ is neglected. The uniform mean pressure( $\bar{p}_0$ ) and temperature( $\bar{T}_0$ ) are the averaged values along the radial direction. The pressure disturbance then can be deduced from Eq. (5) and reads

$$\begin{aligned} \varphi_p(r) = &\frac{\rho_0 c_0^2 \beta \bar{T}_0}{\gamma \bar{p}_0}\varphi_T(r) \\ &- \frac{\rho_0 c_0^2}{\mathrm{i}\gamma He \bar{p}_0 (1-KM)}\left[\frac{\mathrm{d}}{r\mathrm{d}r}\left(r\varphi_r\right) - \frac{\mathrm{i}m}{r}\varphi_\theta - \mathrm{i}HeK\varphi_z\right]. \end{aligned} \quad (10)$$

Substituting the above equation into Eqs. (6) and (7) respectively gives the mathematical formulations of viscothermal linear disturbance dynamics in terms of velocity and temperature perturbations, which reads respectively

$$
\begin{aligned}
&\mathrm{i}\left(1-KM\right)\varphi_r=-\frac{\beta\overline{T}_0}{\gamma He}\left(2\frac{c_0}{\overline{c}_0^2}\frac{\mathrm{d}c_0}{\mathrm{d}r}\varphi_T+\frac{c_0^2}{\overline{c}_0^2}\frac{\mathrm{d}\varphi_T}{\mathrm{d}r}\right)\\
&+\frac{1}{S^2}\begin{bmatrix}\frac{1}{r}\frac{\mathrm{d}}{\mathrm{d}r}\left(r\frac{\mathrm{d}\varphi_r}{\mathrm{d}r}\right)-\frac{\left(m+1\right)^2}{r^2}\varphi_r\\ -He^2K^2\varphi_r+\frac{2m}{r^2}\left(\varphi_r+\mathrm{i}\varphi_\theta\right)\end{bmatrix}\\
&+\frac{1}{S^2}\begin{bmatrix}\frac{A_{\mathrm{Re}}}{\mathrm{i}\gamma He}\frac{c_0^2}{\overline{c}_0^2\left(1-KM\right)}\\ +\left(\frac{\zeta}{\eta}+\frac{1}{3}\right)\end{bmatrix}\frac{\mathrm{d}}{\mathrm{d}r}\begin{bmatrix}\frac{1}{r}\frac{\mathrm{d}}{\mathrm{d}r}\left(r\varphi_r\right)-\frac{\mathrm{i}m}{r}\varphi_\theta\\ -\mathrm{i}HeK\varphi_z\end{bmatrix}\\
&+\frac{1}{\mathrm{i}\gamma He^2}\begin{bmatrix}\frac{2}{\left(1-KM\right)}\frac{c_0\mathrm{d}c_0}{\overline{c}_0^2\mathrm{d}r}\\ +\frac{c_0^2K}{\overline{c}_0^2\left(1-KM\right)^2}\frac{\mathrm{d}M}{\mathrm{d}r}\end{bmatrix}\begin{bmatrix}\frac{1}{r}\frac{\mathrm{d}}{\mathrm{d}r}\left(r\varphi_r\right)-\frac{\mathrm{i}m}{r}\varphi_\theta\\ -\mathrm{i}HeK\varphi_z\end{bmatrix}.
\end{aligned}
\tag{11}
$$

$$
\begin{aligned}
&\mathrm{i}\left(1-KM\right)\varphi_\theta=\frac{\mathrm{i}m\beta\overline{T}_0}{\gamma He}\frac{c_0^2}{\overline{c}_0^2}\frac{\varphi_T}{r}\\
&+\frac{1}{S^2}\begin{bmatrix}\frac{1}{r}\frac{\mathrm{d}}{\mathrm{d}r}\left(r\frac{\mathrm{d}\varphi_\theta}{\mathrm{d}r}\right)-\frac{\left(m+1\right)^2}{r^2}\varphi_\theta\\ -He^2K^2\varphi_\theta+\frac{2m}{r^2}\left(\varphi_\theta-\mathrm{i}\varphi_r\right)\end{bmatrix}\\
&-\frac{\mathrm{i}m}{S^2}\begin{bmatrix}\frac{A_{\mathrm{Re}}}{\mathrm{i}He\gamma}\frac{c_0^2}{\overline{c}_0^2\left(1-KM\right)}\\ +\left(\frac{\zeta}{\eta}+\frac{1}{3}\right)\end{bmatrix}\frac{1}{r}\begin{bmatrix}\frac{1}{r}\frac{\mathrm{d}}{\mathrm{d}r}\left(r\varphi_r\right)-\frac{\mathrm{i}m}{r}\varphi_\theta\\ -\mathrm{i}HeK\varphi_z\end{bmatrix}.
\end{aligned}
\tag{12}
$$

$$
\begin{aligned}
&\mathrm{i}\left(1-KM\right)\varphi_z=\frac{\mathrm{i}\beta\overline{T}_0}{\gamma}\frac{c_0^2}{\overline{c}_0^2}K\varphi_T-\frac{1}{He}\frac{\mathrm{d}M}{\mathrm{d}r}\varphi_r\\
&+\frac{1}{S^2}\left[\frac{1}{r}\frac{\mathrm{d}}{\mathrm{d}r}\left(r\frac{\mathrm{d}\varphi_z}{\mathrm{d}r}\right)-\frac{m^2}{r^2}\varphi_z-He^2K^2\varphi_z\right]\\
&-\mathrm{i}\frac{He}{S^2}\begin{bmatrix}\frac{A_{\mathrm{Re}}}{\mathrm{i}\gamma He}\frac{c_0^2}{\overline{c}_0^2\left(1-KM\right)}\\ +\left(\frac{\zeta}{\eta}+\frac{1}{3}\right)\end{bmatrix}K\begin{bmatrix}\frac{1}{r}\frac{\mathrm{d}}{\mathrm{d}r}\left(r\varphi_r\right)-\frac{\mathrm{i}m}{r}\varphi_\theta\\ -\mathrm{i}HeK\varphi_z\end{bmatrix}.
\end{aligned}
\tag{13}
$$

$$
\begin{aligned}
&\mathrm{i}\left(1-KM\right)\varphi_T=\frac{\gamma\beta\overline{c}_0^2}{S^2c_p}\frac{M}{r}\frac{\mathrm{d}}{\mathrm{d}r}\left(r\frac{\mathrm{d}M}{\mathrm{d}r}\right)\varphi_T\\
&+\frac{\gamma}{\sigma^2S^2}\left[\frac{1}{r}\frac{\mathrm{d}}{\mathrm{d}r}\left(r\frac{\mathrm{d}\varphi_T}{\mathrm{d}r}\right)-\frac{m^2}{r^2}\varphi_T-He^2K^2\varphi_T\right]\\
&-\frac{\beta\overline{c}_0^2}{Hec_p}\frac{T_0c_0^2}{\overline{T}_0\overline{c}_0^2}\left[\frac{1}{r}\frac{\mathrm{d}}{\mathrm{d}r}\left(r\varphi_r\right)-\frac{\mathrm{i}m}{r}\varphi_\theta-\mathrm{i}HeK\varphi_z\right]\\
&-\frac{\gamma}{He}\frac{\mathrm{d}T_0}{\overline{T}_0\mathrm{d}r}\varphi_r+\frac{\gamma\beta\overline{c}_0^2}{S^2c_p}\frac{T_0}{\overline{T}_0}\frac{\mathrm{d}}{r\,\mathrm{d}r}\left(r\frac{\mathrm{d}M}{\mathrm{d}r}\right)\varphi_z\\
&+\frac{2\gamma\overline{c}_0^2}{S^2c_p\overline{T}_0}\frac{\mathrm{d}M}{\mathrm{d}r}\left(\frac{\mathrm{d}\varphi_z}{\mathrm{d}r}-\mathrm{i}HeK\varphi_r\right).
\end{aligned}
\tag{14}
$$

The above expressions use the acoustic Helmholtz number $He=k_0R$, shear wavenumber $S=\sqrt{R^2\omega\rho_0/\eta}$, Prandtl number $\sigma=\sqrt{\eta c_p/\kappa_{th}}$ and acoustic Reynolds number $A_{\mathrm{Re}}=\rho_0R\overline{c}_0/\eta$. In the low frequency disturbance assumption[12-16], the radial velocity dynamics is simplified to $\mathrm{d}\varphi_p/\mathrm{d}r=0$.

As the fluid viscosity is considered, no-slip boundary condition can be obtained, which leads to the vanishing of axial and circumferential velocity disturbances at the wall. Due to the rigid-walled pipeline configuration, the radial velocity disturbance should also vanish at the wall. The thermal conduction of fluid is largely smaller compared with that of solid wall, the temperature disturbance at the wall can be taken zero[14, 23]. As a result, one obtains the following boundary conditions

$$
\varphi_r=\varphi_\theta=\varphi_z=\varphi_T=0,\quad r=1 \tag{15}
$$

In the end, the asymmetric acoustic dynamics under the effects of viscothermal dissipation and axially shear flow convection have been established via coupled equations Eqs.(11)-(14) under the boundary conditions defined by Eq. (15).

If the viscothermal effect is not taken into consideration, the mean temperature (Eq.(4)) should be invariable($T_0=\overline{T}_0$) in the radial direction, leading to the uniform adiabatic sound speed ($c_0=\overline{c}_0$). Furthermore, the disturbance should be isentropic ($s'=0$) and a simpler mathematical equation governing the isentropic disturbance should be obtained and read

$$
\begin{aligned}
&\frac{\mathrm{d}^2\varphi_p}{\mathrm{d}r^2}+\left[\frac{1}{r}+\frac{2K}{\left(1-KM\right)}\frac{\mathrm{d}M}{\mathrm{d}r}\right]\frac{\mathrm{d}\varphi_p}{\mathrm{d}r}-\frac{m^2}{r^2}\varphi_p\\
&+He^2\left[\left(1-KM\right)^2-K^2\right]\varphi_p=0.
\end{aligned}
\tag{16}
$$

A comprehensive deduction of the above equation can be found in Mungur and Plumblee [6] and Chen, et al. [9]. Furthermore, the rigid-walled no-invasion boundary condition($\varphi_r\left(r\right)=0$ at $r=1$) of Eq. (16) can be simplified to $\mathrm{d}\varphi_p/\mathrm{d}r\big|_{r=1}=0$.

*B. Iterative calculation method*

Physically speaking, disturbances of velocity, temperature as well as pressure in Eqs. (11)-(14) and (16)) should be continuous and finite in a circular cylindrical pipeline. They can be represented as Fourier sequences of orthogonal and complete functions which are selected on the users' convenience. Under the constraint of circular cylindrical coordinate, Bessel functions may be a good choice and have been used in the authors' previous study on axisymmetric disturbance dynamics[20, 21]. In the case of finite asymmetric disturbances along the range $r\in\left[0,1\right]$, one can obtain the following expressions of velocity and temperature disturbances

$$
\begin{aligned}
&\varphi_r\left(r\right)=\sum_{j=1}^{\infty}C_j^rJ_{m+1}\left(\lambda_j^rr\right),\ \varphi_\theta\left(r\right)=\sum_{j=1}^{\infty}C_j^\theta J_{m+1}\left(\lambda_j^\theta r\right),\\
&\varphi_z\left(r\right)=\sum_{j=1}^{\infty}C_j^zJ_m\left(\lambda_j^zr\right),\ \varphi_T\left(r\right)=\sum_{j=1}^{\infty}C_j^TJ_m\left(\lambda_j^Tr\right),
\end{aligned}
\tag{17}
$$

where $J_m\left(\cdot\right)$ denotes the $m$th-order Bessel function of the first kind and $m$ corresponds to the circumferential index. $C_j^r$, $C_j^\theta$, $C_j^z$ and $C_j^T$ are coefficients and do not vanish simultaneously if linear disturbances exist. The orthogonality and completeness can be guaranteed by mathematical results from substitution of Eq. (17) into Eq. (15), which read

$$J_{m+1}\left(\lambda_j^r\right)=J_{m+1}\left(\lambda_j^\theta\right)=J_m\left(\lambda_j^z\right)=J_m\left(\lambda_j^T\right)=0. \quad (18)$$

Using the same procedure, the corresponding Bessel-Fourier function of Eq. (16) should be

$$\varphi_p(r)=\sum_{j=1}^{\infty}C_j^p J_m\left(\lambda_j^p r\right),\ J_{m-1}\left(\lambda_j^p\right)-J_{m+1}\left(\lambda_j^p\right)=0. \quad (19)$$

It should be noticed that present paper deals with the rigid-walled boundary condition, where the acoustic impedance is infinite. On the other hand, the acoustic impedance of non-rigid pipeline wall should be considered, which has been massively studied in the inviscid flow. Mathematical models can be found in the work of Brambley [24], Richter, et al. [25], Vilenski and Rienstra [26]. Under the consideration of viscothermal dissipation, the viscous effect leads to no-slip boundary condition for both mean flow and disturbance. Under the vanishing temperature disturbance at the wall, mathematical modelling of acoustic impedance can be found in the authors' previous work[27].

Substituting Eq. (17) into Eqs. (11)-(14) and doing integration procedures as shown in the supporting document, Eqs. (11)-(14) can be deduced into linear equations

$$\mathbf{G}(K,M,R,\omega,m)\mathbf{X}=0, \quad (20)$$

The vector $\mathbf{X}$ denotes the coefficients of Fourier-Bessel sequences and reads

$$\mathbf{X}=\left[C_1^r,\cdots,C_N^r,C_1^\theta,\cdots,C_N^\theta,C_1^z,\cdots,C_N^z,C_1^T,\cdots,C_N^T\right]^{\mathrm{T}},$$

where superscript 'T' represents the vector transposition and $\mathbf{G}$ is a $4N\times4N$ matrix whose elements are functions of the dimensionless axial wavenumber $K$, Mach number $M$, pipeline radius $R$ and angular frequency $\omega$, which constitutes the acoustic Helmholtz number and shear wavenumber in the governing equations.

On the existence of linear disturbances, the coefficient vector satisfies $\mathbf{X}\neq\mathbf{0}$, indicating that nonzero solutions to Eq. (20) exit. Therefore, the corresponding determinant vanishes with

$$\det\left[\mathbf{G}(K,M,R,\omega,m)\right]=0. \quad (21)$$

If the local Mach number, pipeline radius and angular frequency are given, the dimensionless axial wavenumber can be solved iteratively[20, 21]. In the above equation, the number of Bessel functions must be infinite to satisfy the completeness of Eq. (17). Due to the computing limitation, a finite number is preferred, which will lead to truncation errors. A proper way is to numerically verify the calculation convergence under different values of $N$. Such treatment was reported in Galerkin method[28]. The convergence study for axisymmetric viscothermal acoustic disturbance can be found in Chen, et al. [17] and Chen, et al. [20]

## III. NUMERICAL STUDY

This section deals with numerical analysis of the relative phase velocity ($1/K_{\mathrm{R}}$ with $K_{\mathrm{R}}$ being the real component of the dimensionless axial wave number) and attenuation coefficient ($A=\left|8.686k_0K_{\mathrm{I}}\right|:\mathrm{dB/m}$ with $K_{\mathrm{I}}$ being the imaginary component of the dimensionless axial wave number) of linear disturbances under the effects of viscothermal dissipation and shear flow convection. Furthermore, high frequency acoustic flow measurement performance is given. Comparison with isentropic acoustic features is carefully addressed. In the numerical study, pure water at temperature $20\,^0\mathrm{C}$ is adopted with its parameters listed as follows: $\beta=0.207\times10^{-3}\ /^0\mathrm{C}$, $\rho_0=1000\,\mathrm{kg/m^3}$, $T_{\mathrm{wall}}^0=293\mathrm{K}$, $\gamma=1$, $\eta=1\times10^{-3}\,\mathrm{kg/(s\cdot m)}$, $\zeta=2.4\eta$, $\kappa_{\mathrm{th}}=0.5984\,\mathrm{W/(K\cdot m)}$, $c_p=4181.3\mathrm{J/(Kg\cdot K)}$. The pipeline radius is $R=4\mathrm{mm}$. Then, the adiabatic sound speed can be defined as a function of the mean temperature[29]

$$c_0=\sum_{n=0}^{3}\alpha_n\left(T_0-273\right)^n. \quad (22)$$

The corresponding coefficients are $\alpha_0=1449.2$, $\alpha_1=4.6$, $\alpha_2=-0.055$ and $\alpha_3=0.00029$. In the numerical study, only acoustic modes are calculated while hydrodynamic modes are taken out of consideration. The acoustic modes are denoted by $k_{mn}$ with m and n represent the circumferential and axial index respectively. Obviously, $m=0$ stands for the axisymmetric modes.

### A. Acoustic frequency

In this subsection, concentration is imposed on the frequency effect on the relative phase velocity between viscothermal and isentropic acoustic dynamics, which refers to the effect of viscothermal dissipation. To get rid of flow convection, the Mach number is set zero, the isentropic disturbance in the inviscid flow (Eq. (16)) then has a simple analytic solution $\varphi_p=J_m\left(He\sqrt{1-K^2}r\right)$. Under the rigid-walled boundary condition(Eq. (19)), one obtains

$$J_{m-1}\left(He\sqrt{1-K^2}\right)-J_{m+1}\left(He\sqrt{1-K^2}\right)=0. \quad (23)$$

It can be easily solved using standard iterative numerical methods.

In Fig. 1a, comparisons of the relative phase velocity between isentropic(denoted by 'inv' in the legend) and viscothermal asymmetric acoustic modes under the same axial index(n=0) are given. Comparisons of the relative phase velocity among different axial indices are given in Fig. 1b under the same circumferential index (m=1). According to Eq. (23), it can be learned that zero is a common solution of $J_{m-1}(x)-J_{m+1}(x)=0$ under the case of $m\geq1$, which indicates that the corresponding relative phase velocity is 1. It physically refers to the plane wave. To avoid confusions in the following analysis, the authors do not take the solution $K=1$ for asymmetric acoustic dynamics ($m\geq1$).

As shown in Fig. 1a, only the plane mode (m= n=0) can exist and is independent on the acoustic frequency, which is similar to the literature work[18]. For a specific acoustic mode, it has individual cut-on frequency. Specifically, the cut-on frequency of acoustic mode with higher circumferential or axial index is larger as shown in Figs. 1a and 1b. More importantly, Fig 1 shows that the cut-on frequency of isentropic acoustic mode in inviscid fluid is larger than that of non-isentropic acoustic mode in viscothermal fluid. The phenomenon shows that the viscothermal dissipation obviously affects the features of acoustic dynamics.

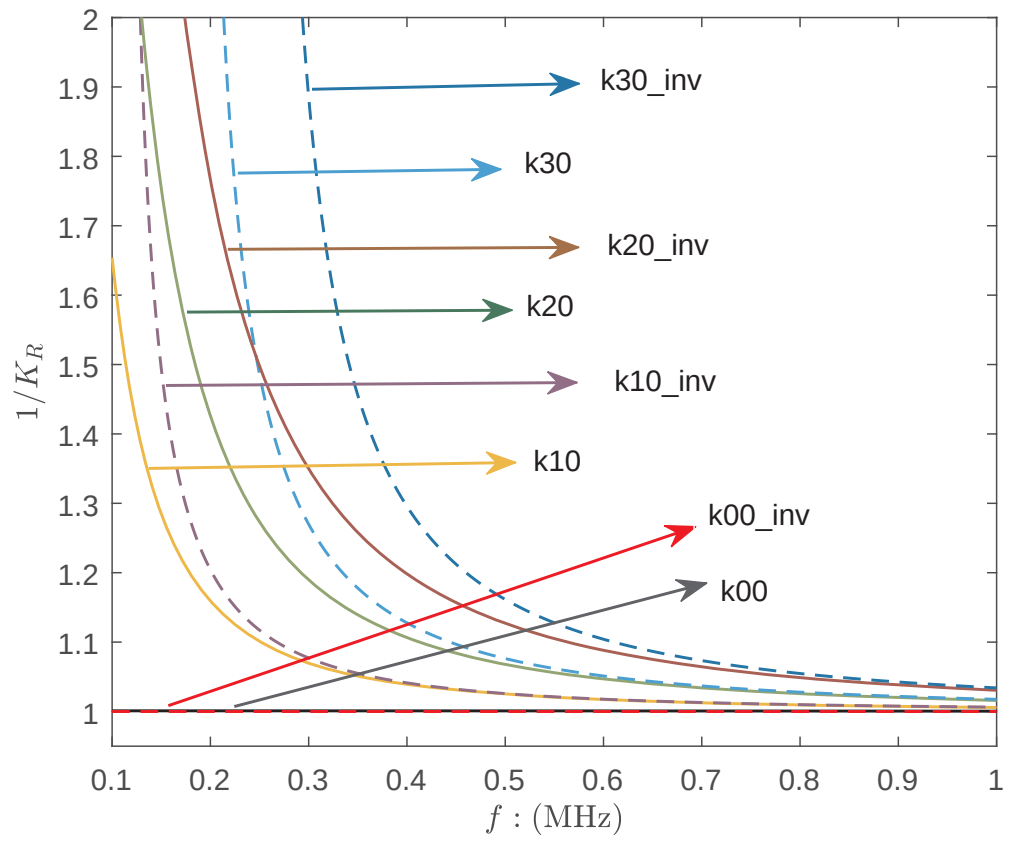

a

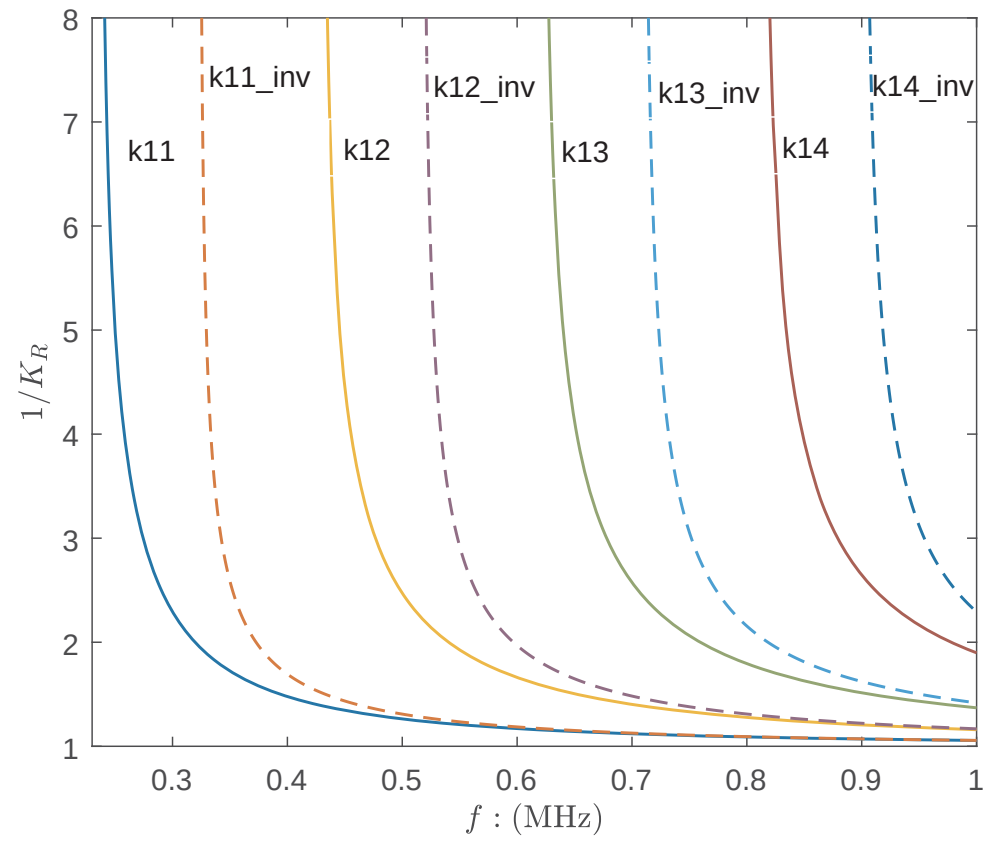

b

Fig. 1(Color online). Comparisons of the relative phase velocity as function of acoustic frequency between inviscid and viscothermal fluid models. a: the relative phase velocity of acoustic modes with different circumferential indices and zeroth axial index(n=0). b: the relative phase velocity of acoustic modes with different axial indices and first circumferential index(m=1).

In the case of isentropic disturbance propagating in the inviscid fluid, the corresponding cut-on frequency of specific acoustic mode($k_{mn}$) can be calculated through

$$\omega = j_{mn}\bar{c}_0 / R, \tag{24}$$

where $j_{mn}$ is the nth solution of $J_{m-1}(x) - J_{m+1}(x) = 0$. It should be noticed that Eq. (24) is obtained by solving Eq. (23) under the regulation of $K = 0$. On the other hand, calculating the cut-on frequency of a non-isentropic acoustic mode affected by viscothermal dissipation is much complex. Mathematically, obtaining a simple analytical expression like Eq. (24) is difficult. On the other hand, one can get the number of acoustic cut-on modes by drawing Eq. (21) as a function of $K$ if the specific frequency is given. An example can be found in Chen, et al. [30].

## *B. Flow Mach number*

This subsection deals with the effect of flow Mach number on the relative phase velocity and attenuation coefficient of asymmetric acoustic modes. In the numerical analysis, the flow profile is assumed parabolic $M(r) = 2\bar{M}(1-r^2)$, where $\bar{M}$ is mean Mach number. The acoustic frequency is taken $f = 1\,\text{MHz}$. Fig. 2 compares the relative phase velocity of asymmetric modes (k10, k11, k12, 13) between inviscid and viscothermal fluid configurations. Specifically, Figs. 2a and 2b show the relative phase velocity propagating in the downstream and upstream directions respectively.

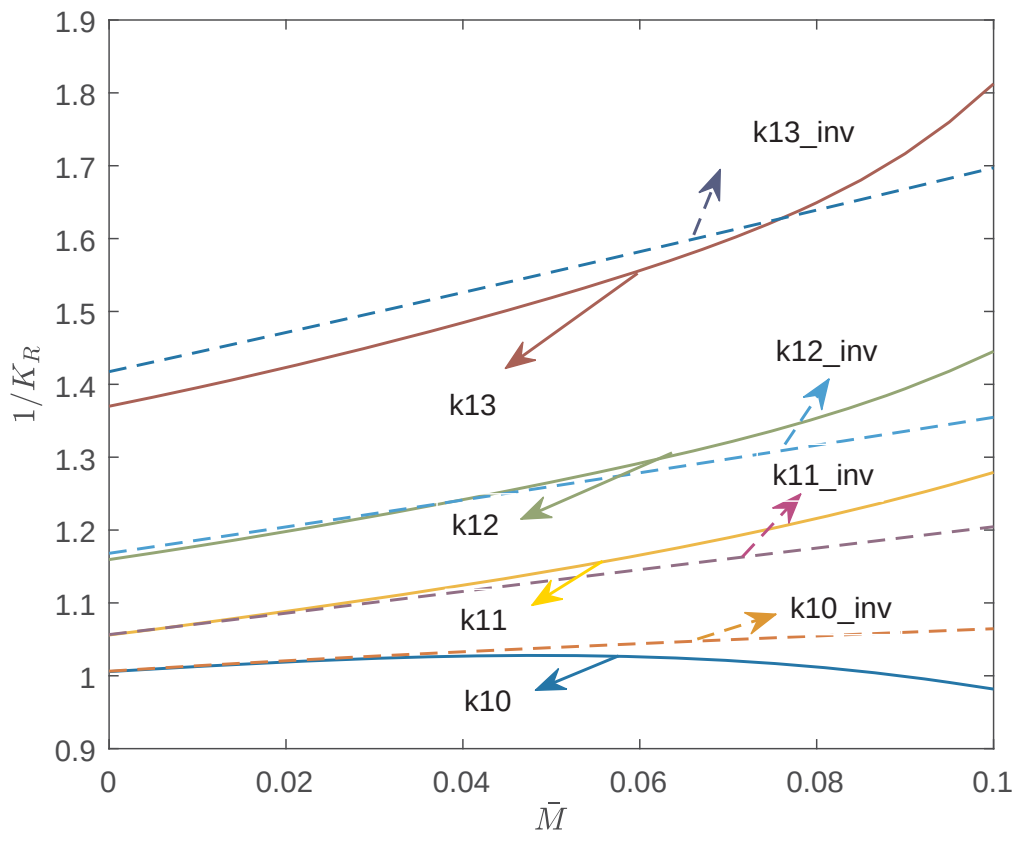

a

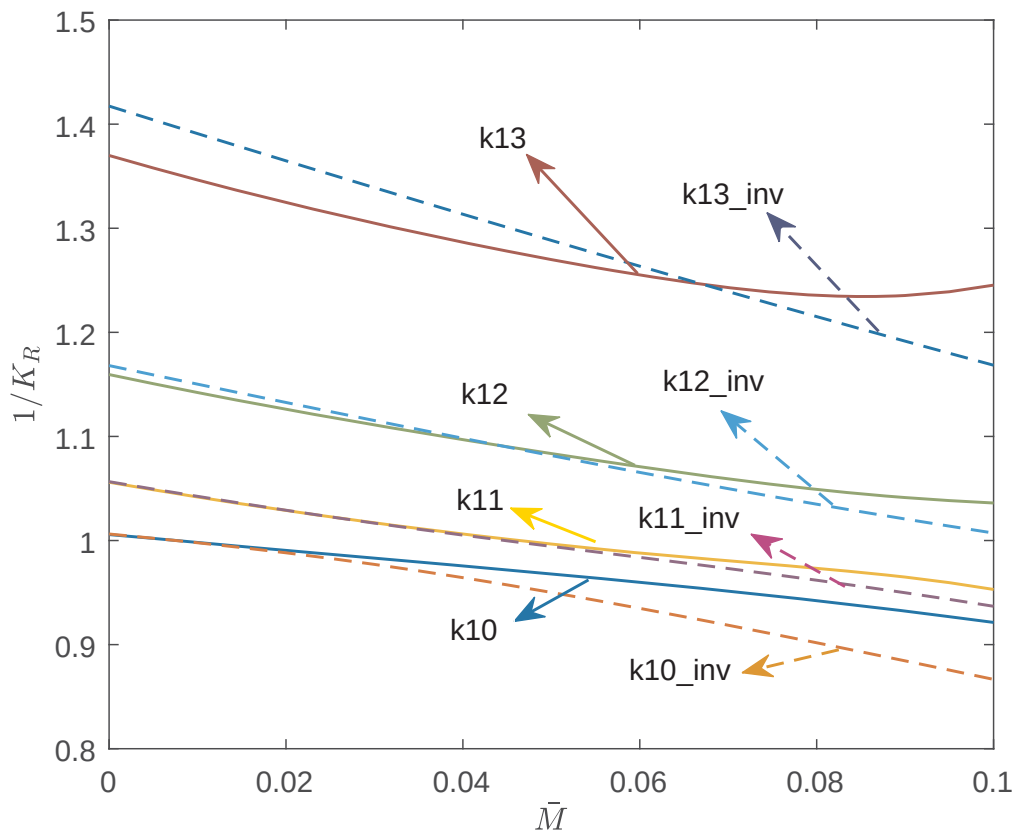

b

Fig. 2(color online). Comparisons of the relative phase velocities as functions of mean Mach number between inviscid and viscothermal fluid models. a: relative phase velocities of acoustic modes propagating in the downstream direction with different axial indices and first circumferential index. b: relative phase velocities of acoustic modes propagating in the upstream direction with different axial indices and first circumferential index.

In the downstream propagation as shown in Fig. 2a, the relative phase velocity of acoustic mode with zeroth axial index(k10 and k10_inv) changes differently from that of other acoustic modes. Specifically, the relative phase velocity of acoustic mode with zeroth axial index increases along with a small mean Mach number. Meanwhile, the absolute gradient value of flow profile ($\mathrm{d}M(r)/\mathrm{d}r = 4\bar{M}r$) increases, which will affect the acoustic dynamics. Its effect will slow down the relative phase velocity of acoustic mode with zeroth axial index and speed up the relative phase velocity of acoustic mode with higher axial index( $n \geq 1$ ) in the downstream propagation. In the literature, this phenomenon was reported in the analysis of axisymmetric isentropic acoustic dynamics[6] and axisymmetric viscothermal acoustic dynamics[29]. Fig. 2a shows that similar tendency can be found in the asymmetric acoustic dynamics. Compared between inviscid and viscothermal fluid configurations, one observes that the viscothermal dissipation enlarges the tendency. Specifically, the viscothermal dissipation slows down the relative phase velocity of non-isentropic acoustic mode with zeroth axial index and speeds up the relative phase velocity of non-isentropic acoustic mode with higher axial index. Furthermore, the enlargement becomes more apparent in the case of a higher axial index.

In the upstream propagation as shown in Fig. 2b, the relative phase velocity decreases against the Mach number due to flow convection. With the increase of flow gradient( $\mathrm{d}M\left(r\right)/\mathrm{d}r = 4\bar{M}r$ ), the shear effect strengthens the flow convection on the acoustic mode with zeroth axial index('k10_inv'). On the other hand, the shear effect on other acoustic modes is not obvious. However, the viscothermal dissipation increases the relative phase velocity of all the non-isentropic acoustic modes. With larger axial index, the effect of viscothermal dissipation becomes more apparent.

Fig. 3 shows the attenuation coefficients of asymmetric acoustic modes propagating in the downstream and upstream directions due to viscothermal dissipation. As no energy absorption occurs at the rigid wall, the isentropic acoustic modes propagating in the inviscid fluid confront zero attenuation. In the previous research[30], the authors analyzed the attenuation coefficients of the first four axisymmetric acoustic modes(m=0 and n = 0,1,2,3) propagating in the uniform flow due to viscothermal dissipation. It was shown that the uniform flow convection enlarges the attenuation coefficient of acoustic mode propagating in the upstream direction and decreases the attenuation coefficient in the downstream direction. Furthermore, acoustic mode with higher axial index(n = 1, 2, 3), except the zero axial index, has larger attenuation coefficient in both downstream and upstream propagations. The above mentioned tendencies can been found in Fig. 3.

Despite the flow convection in the uniform flow, Fig. 3 also reflects the shear regulation on the acoustic modes. Specifically, the shear effect increases the attenuation coefficient of acoustic mode(k10) propagating in the downstream direction. On the other hand, the attenuation coefficient of acoustic mode(k10) propagating in the upstream direction decreases when the shear regulation is larger than the flow convection. With the increase of Mach number, the attenuation coefficient finally increase showing that flow convection become dominant. Fig. 3 also reveals that the shear effect and flow convection interacts nonlinearly(see the parameters of $\nabla \cdot \mathbf{v}'$ in Eqs. (11)-(14)) on acoustic propagation when the Mach number is relatively large.

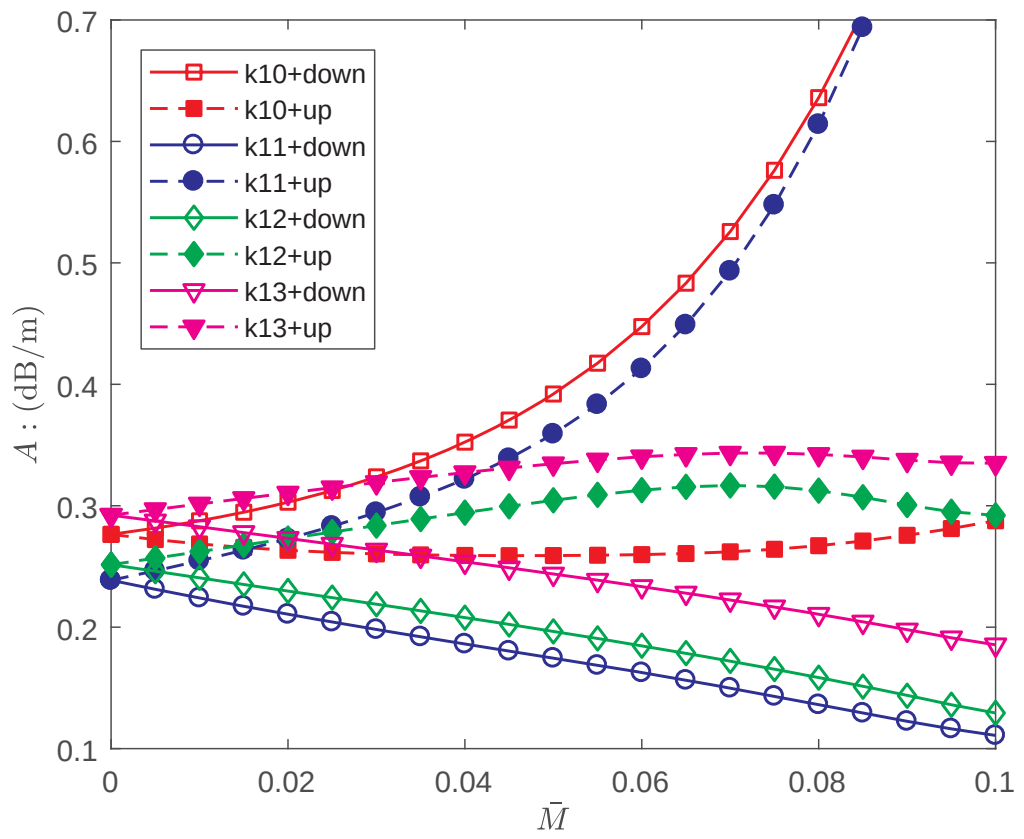

Fig. 3(color online). Comparisons of the attenuation coefficients as functions of mean Mach number of viscothermal acoustic modes propagating in the upstream and downstream directions. The circumferential index is 1 while the axial index changes.

## C. *Measurement performance*

This subsection deals with measurement performance of high frequency acoustic flow measurement using transit-time difference method. In the literature, this problem has been comprehensively analyzed in the case of axisymmetric modes when a laminar flow(its flow profile is parabolic) is present. Readers can consult the work of Lechner [3], Willatzen[4, 5] and Chen, et al.[9, 21, 22]. On the other hand, Chen, et al. [9] analyzed measurement performance of asymmetric isentropic acoustic modes in the inviscid fluid. To the knowledge of the authors, measurement performance using asymmetric acoustic modes under the viscothermal effect has not been comprehensively addressed. Using the measurement deviation by Willatzen[5], which reads

$$E_{\mathrm{mn}} = \frac{\left(K_{\mathrm{R}}^{\mathrm{up}}\right)^{\mathrm{mn}} - \left(K_{\mathrm{R}}^{\mathrm{down}}\right)^{\mathrm{mn}}}{2\bar{M}} - 1, \qquad (25)$$

Fig. 4 shows the measurement performance of asymmetric acoustic modes in the parabolic flow profile, which is a theoretical model of laminar flow. Viscothermal effect is comprehensively analyzed by comparing between the isentropic and non-isentropic acoustic modes.

Fig. 4a compares the difference of measurement deviation between inviscid and viscothermal fluid for asymmetric acoustic modes with the zeroth axial index (k10 and k40). With the increase of Mach number, the absolute measurement deviation in the inviscid fluid model decreases while the absolute measurement deviation in the viscothermal fluid model increases for each mode. The discrepancy shows that viscothermal dissipation is an important factor in the application of acoustic flow measurement. According to Fig. 2, it can be learned that the relative phase velocity of acoustic mode under viscothermal effect is obviously different from that when the inviscid fluid model is adopted. As the measurement deviation (Eq. (25)) is deduced from the relative phase velocity, the distinction can be perceived.

Figs. 4b-4c shows the difference of measurement deviation of asymmetric acoustic modes with the same circumferential

index but different axial indices. It can be learned that the difference of measurement deviation between the inviscid and viscothermal fluid models is extremely distinct. Furthermore, the variation tendency of measurement deviation with lower axial index is more sensitive to the shear effect as the deviation declines with small Mach number but goes up with large Mach number. The viscothermal dissipation enhances the tendency as shown by lines of k11, k12, k41, and k42 in Figs. 4b and 4c. The performance deviation of acoustic modes with larger axial index(k13 and k43) is more tolerant to shear effect.

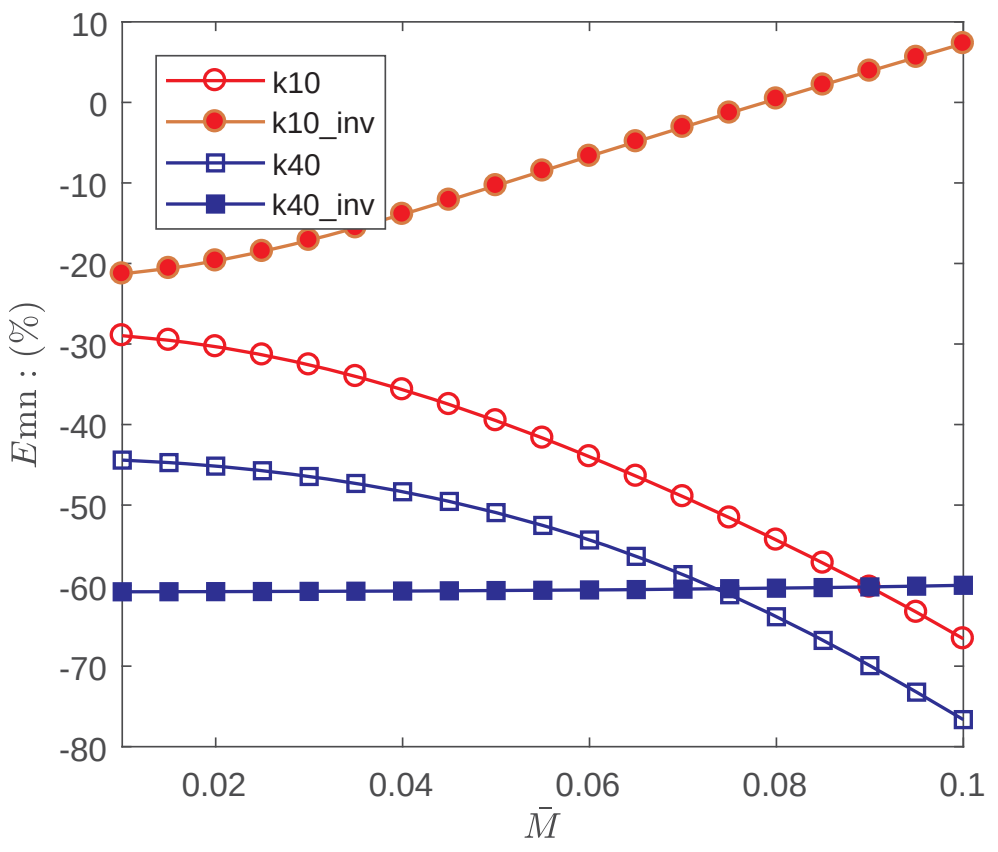

a

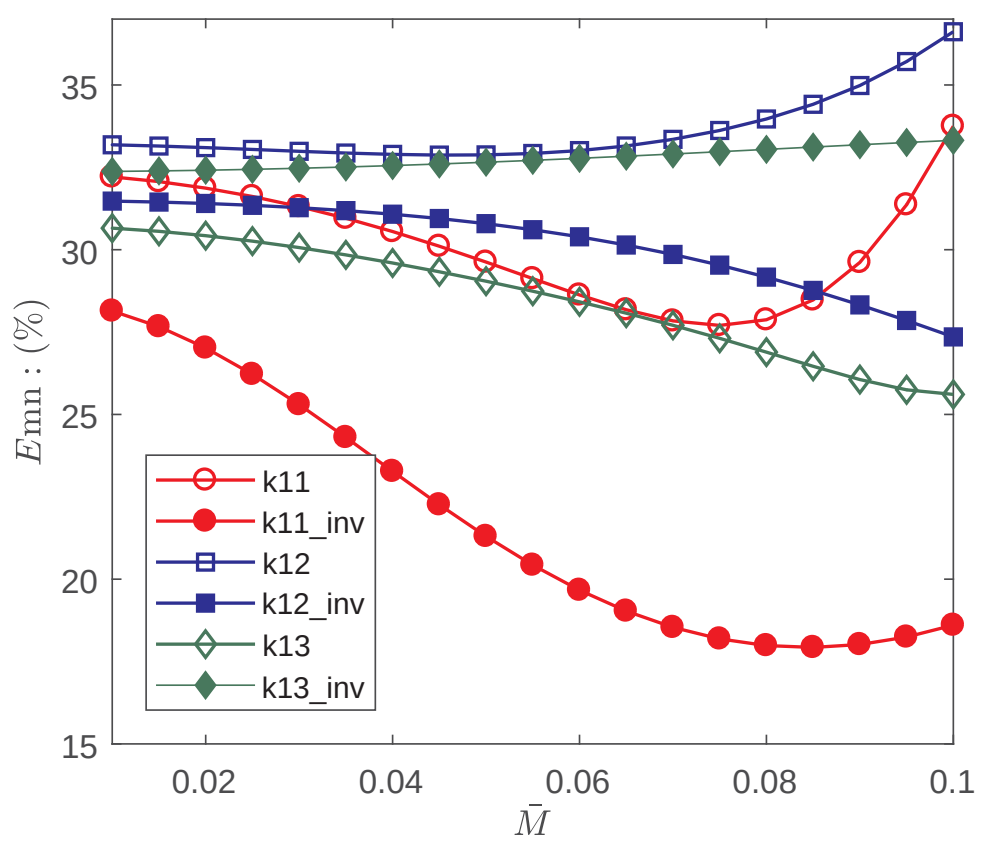

b

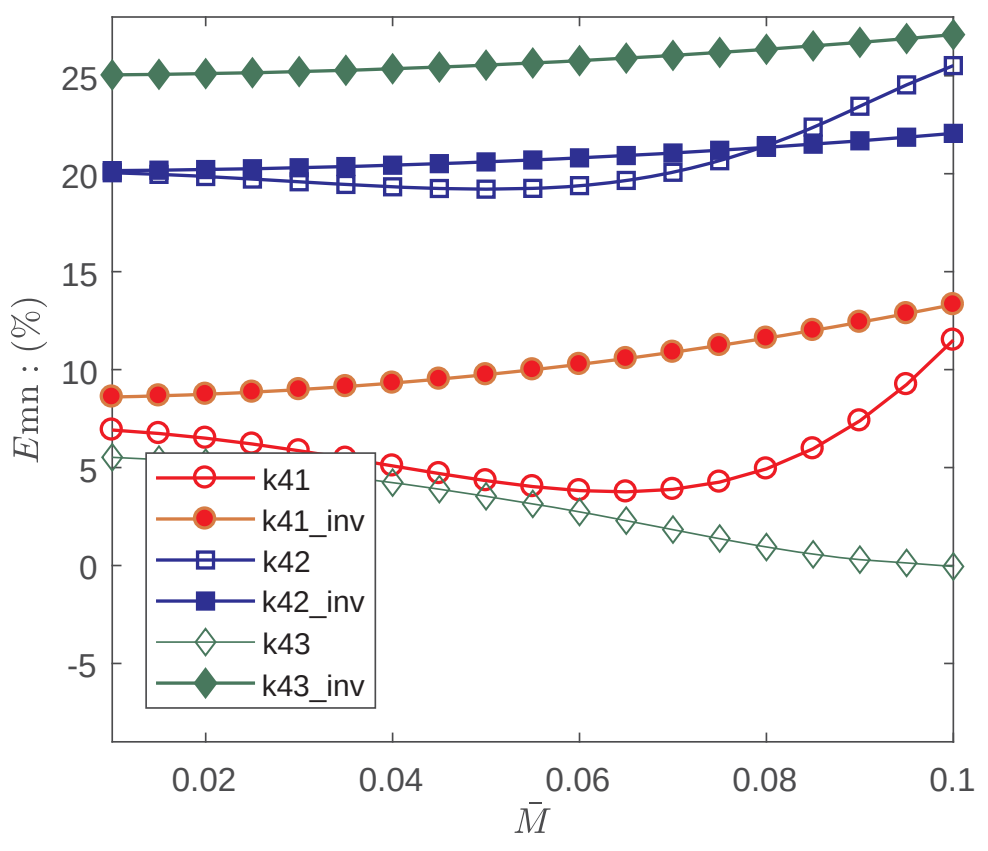

c

Fig. 4. Comparisons of measurement deviations of asymmetric acoustic modes between inviscid and viscothermal fluid models. a: asymmetric wave modes with the same axial index but different circumferential indices. b: asymmetric wave modes with the same circumferential index but different axial indices.

## IV. CONCLUSION

Present paper addresses mathematical formulation of asymmetric acoustic dynamics under the effects of viscothermal dissipation and shear flow convection as many asymmetric modes are cut-on in the application of high frequency acoustic flow measurement where acoustic energy dissipation should be taken into consideration. Mathematical deduction of asymmetric linear disturbance is given for rigid-walled shear pipeline flow where viscothermal effect is considered. An iterative numerical calculation procedure, which is similar to Galerkin method, is presented to solve the established mathematical model.

Parametric analysis of frequency and parabolic flow Mach number on the relative phase velocity and attenuation coefficient are given. Comparing with inviscid fluid shows that viscothermal dissipation alters the cut-on frequency of any non-plane mode. The shear effect and viscothermal dissipation interact nonlinearly in the case of high Mach number and complicate the tendency of acoustic dynamics(the relative phase velocity and attenuation coefficient).

Theoretical comparisons of acoustic flow measurement using asymmetric modes between inviscid and viscothermal fluid models are literally given. It presents obvious distinction of measurement performance between inviscid and viscothermal fluid models, which indicates that ignorance of viscothermal dissipation in the high frequency acoustic flow measurement may lead to unacceptable measurement errors. In the case of high Mach number, the nonlinear interaction between shear convection and viscothermal dissipation complicates the measurement performance. Furthermore, different acoustic mode has its own measurement deviation, which shows that strictly distinguishing the acoustic modes is required in the application of high frequency acoustic flow measurement. Fortunately, such task can be settled due to different inherent phase velocity of acoustic mode shown in Figs. 1 and 2. With the distinguished acoustic mode, theoretical compensation can be given according to Fig. 4.

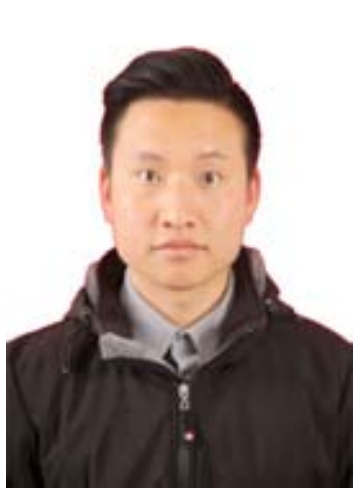

**Yong Chen** was borne in Leshan, Sichuan, PRC, on June 28, 1985. He received B.S. degree from Sichuan University in 2007. He received M.S. and Ph. D. degrees in 2009 and 2013 at National University of Defense Technology(NUDT). Until 2013, he serves as a teacher at NUDT. He is now an academic visitor at Department of Applied Mathematics and Theoretical Physics, University of Cambridge. His research interests include theoretical acoustics and fluid mechanism.

He gets funding from National Natural Science Foundation of China, Lift engineering young talent of China Association for Science and Technology and Yong top-notch talent of NUDT.

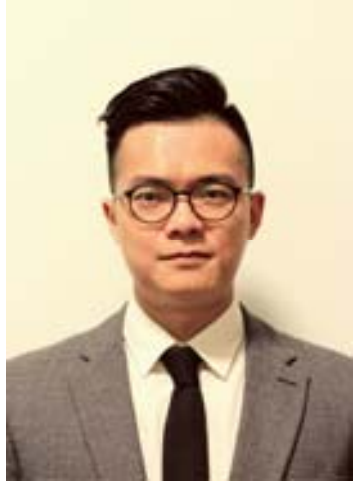

**Bo Yuan** was borne in Xiaogan, Hubei, PRC, on May 11, 1985. He received B.S. and Ph. D. degree from NUDT in 2007 and 2013. From 2013 to 2018, he serves as a lecturer at Army Logistics University of PLA. His research interests include underwater acoustics and exoskeleton.

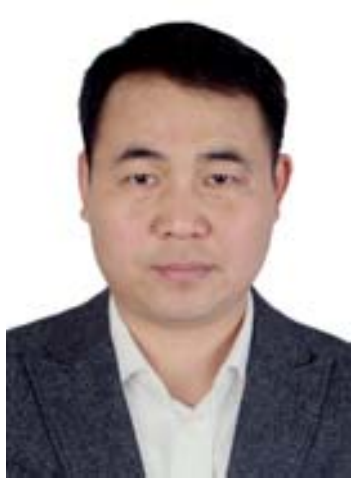

**Xiaoqian Chen** was borne in Shuangfeng, Hunan, PRC, on September 23, 1975. He received B.S. and Ph.D. degrees from NUDT. After graduation, He serves as a teacher in NUDT. From 2006, He becomes a professor in NUDT. His research interests include fluid mechanics, multidisciplinary design optimization.

He was awarded the “Qiu Shi Outstanding Youth Prize for Practical Engineering” and the “Science & Technology Award for Chinese Youth” by China Association for Science and Technology in 2016. In 2017, he was selected to be a winner of the National Science Fund for Distinguished Young Scholars.

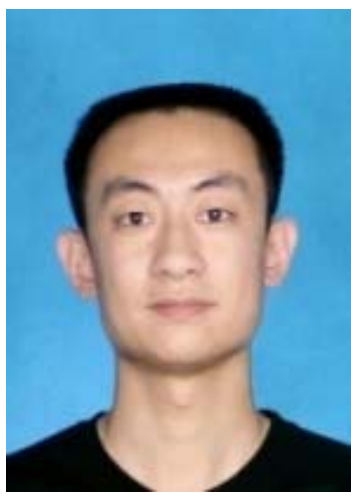

**Lei Qi** was borne in Weifang, Shandong, PRC, on October 22, 1985. He received B.S. and M.S. degrees in 2009 and 2012 from Tianjin University. From 2012 to 2014, he was an assistant engineer at Beijing Institute of Spacecraft Environment Engineering. From 2014, He serves as an engineer at Beijing Institute of Spacecraft Environment Engineering. He is studying for a Ph.D. degree at State Key Laboratory of Precision Measurement Technology and Instrument, Tianjin University. He researches in leak detection and acoustic emission detection.